\documentclass[5p]{elsarticle}

\usepackage{graphicx}

\usepackage{amssymb}

\journal{TIPP09 Proceedings in NIMA}

\begin{document}

\begin{frontmatter}



\title{An imaging time-of-propagation system for charged particle identification at a super B factory}

%
%
%
%
%
\author[First]{K.~Nishimura\corref{cor1}}
\ead{kurtisn@phys.hawaii.edu}
\cortext[cor1]{Corresponding author. Tel.: +1-808-956-2942; fax: +1-808-956-2930.} 
\author[First]{T.~Browder}
\author[First]{H.~Hoedlmoser}
\author[First]{B.~Jacobson}
\author[First]{J.~Kennedy}
\author[First]{M.~Rosen}
\author[First]{L.~Ruckman}
\author[First]{G.~Varner}
\author[First]{A.~Wong}
\author[First]{W.~Yen}

%
%
%
\address[First]     {Department of Physics and Astronomy, University of Hawaii, 2505 Correa Road, Honolulu HI 96822, USA}

%
%
%
%
%
\begin{abstract}

Super B factories that will further probe the flavor sector of the Standard Model and physics
beyond will demand excellent charged particle identification (PID), particularly $K/\pi$ separation, for
momenta up to 4 GeV/c, as well as the ability to operate under beam backgrounds significantly
higher than current B factory experiments. We describe an Imaging Time-of-Propagation (iTOP)
detector which shows significant potential to meet these requirements. Photons emitted from charged
particle interactions in a Cerenkov radiator bar are internally reflected to the end of the bar, where
they are collected on a compact image plane using photodetectors with fine spatial segmentation
in two dimensions. Precision measurements of photon arrival time are used to enhance the two
dimensional imaging, allowing the system to provide excellent PID capabilities within a reduced
detector envelope. Results of the ongoing optimization of the geometric and physical properties of
such a detector are presented, as well as simulated PID performance. Validation of simulations is
being performed using a prototype in a cosmic ray test stand at the University of Hawaii.

\end{abstract}

%
%
%
%
%
%

\end{frontmatter}


%
%
%
%
%
%

\section{Introduction}

Particle identification (PID) has played a key role in the success of the Belle and BaBar B factory experiments, and will continue to be essential at a next generation Super B Factory.  In particular, discrimination of kaons and pions is vital for reconstruction of B meson decays as well as flavor tagging used to extract measurements of time dependent CP asymmetries \cite{ref1}.  PID schemes that focus an internally reflected Cerenkov ring onto a large two dimensional imaging plane have demonstrated excellent PID capability \cite{ref2}, but with prohibitive space requirements.  Other schemes that measure the time-of-propagation and only one spatial dimension of the Cerenkov ring can operate with a smaller detector footprint, but at the cost of reduced PID performance \cite{ref3,ref4}.  The tradeoff between performance and space motivates a compromise between the two concepts, where precision timing measurements are coupled with a compact two dimensional image plane to provide PID capabilities enhanced over a single spatial dimensional readout scheme, but with much smaller space requirements than a full focusing DIRC.  We describe here the details of such an Imaging Time-of-Propagation detector (iTOP) suitable for the barrel region of a super B factory, including results of ongoing detector optimization studies and simulated performance results.

%
%
%
%
%
\begin{figure}[hbt]
\begin{center}
\includegraphics*[scale=0.65]{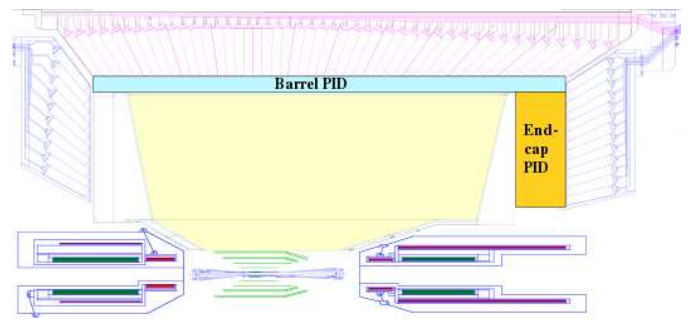}
\end{center}
\caption{\label{fig:fig0}
A partial cross section of the proposed KEKB detector upgrade, showing the regions allotted for barrel and endcap PID devices.  The length of the rectangle denoted as barrel PID is just under 3 m long.
}
\end{figure}

\section{Design criteria and iTOP concept}

\begin{figure}[hbt]
\begin{center}
\includegraphics*[scale=1.0]{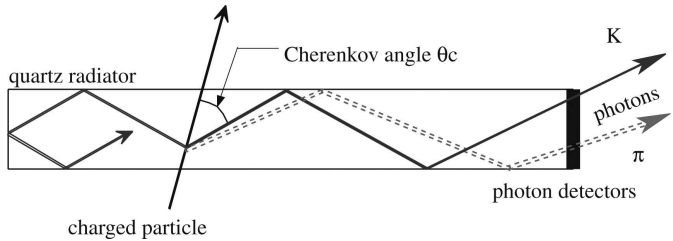}
\end{center}
\caption{\label{fig:fig0a}
A side view schematic of the TOP concept \cite{ref5}.  The path length difference of Cerenkov photons emitted from a pion and kaon within the radiator bar results in a difference in arrival time at the photodetector.
}
\end{figure}

Space places restrictive limits on the geometry of a barrel PID device.  For example, the PID system at the proposed KEKB detector upgrade must fit within the approximate envelope of the existing Belle time-of-flight detector, as shown in Figure \ref{fig:fig0}.  Furthermore, the amount of material comprising the device must be kept to a minimum to avoid degradation of electromagnetic calorimeter performance.  One way to meet the compactness requirement is to place the photodetectors directly at the end of the radiator bar, a concept known as the time-of-propagation counter (TOP).  A basic schematic of the TOP concept is shown in Figure \ref{fig:fig0a}.  Such a scheme may also involve a focusing mirror on one end of the radiator, which helps to reduce chromatic dispersion on the measured timings.  Studies of such a device are detailed elsewhere \cite{ref5}, but in general this approach utilizes high precision timing with relatively fine pixelization in one of the two spatial dimensions.  The other dimension typically has very coarse or unit pixelization.  The timing for a TOP detector is particularly crucial as the ambiguities of possible photon momenta cannot be resolved from the photon detection position alone.  Furthermore, the measured photon arrival time includes the photon propagation time as well as the charged particle time-of-flight from the interaction point.  In principle the time-of-flight information provides some extra $K/\pi$ discrimination, but this feature may render the TOP vulnerable to uncertainties in the global event starting time.

%
%
%
%
%

The iTOP concept builds upon the basic TOP configuration by adding a region of vertical expansion before the detection plane.  Although similar in concept to the stand-off box used in other DIRC designs \cite{ref9}, the size of the this expansion region is reduced from the meter to centimeter scale.  By forming the image further from the bar end, the ambiguities in measuring the Cerenkov angle are reduced, and partially decoupled from the timing measurements.  The expected benefits of this addition must be determined for a given photodetector pixel size, efficiency, and timing resolution, but qualitatively the design benefits from a redundant measurement of the Cerenkov photon parameters as well as a reduced background rate per unit image plane area.  As with the basic TOP, the addition of a focusing mirror at one end of the radiator bar can be considered to improve image quality and mitigate degradation due to chromatic dispersion.

%
%
%
%
%
\begin{figure}[hbt]
\begin{center}
\includegraphics*[scale=0.25]{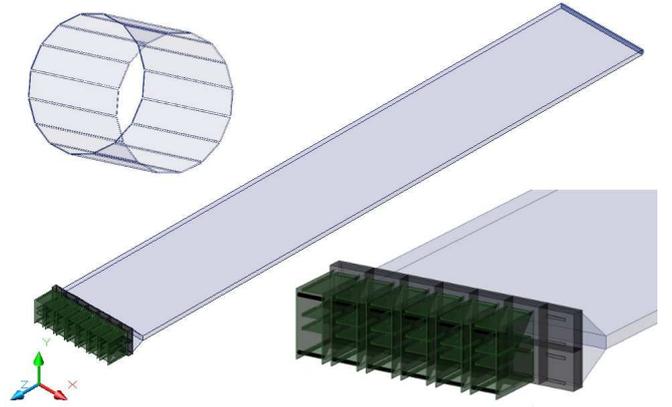}
\end{center}
\caption{\label{fig:fig1}
The center shows an example of a single iTOP unit (approximately 2.8 m in total length) suitable to be arrayed into a barrel PID device (upper left inset, no support structure shown).  The forward end of each bar contains a focusing mirror.  In the backward direction, a readout plane of 10 cm height (enlarged with photodetectors and readout electronics in lower right inset) is separated from the bar end by an expansion region of roughly 4 cm length.
}
\end{figure}

\section{Initial and current designs}

The initial iTOP design utilized readouts on both ends of the radiator bar, with standoff regions constructed of a high index optical glass.  The transition from the fused silica radiator to the high index glass compresses the image into a smaller space, with the intent of collecting the image photons using a highly pixelated solid state readout.  This design was found to be impractical due to difficulties in the transmission through the final standoff region into the photodetectors, as well as performance limitations of the photodetectors themselves.

Current designs utilize a standoff region made out of the same material as the radiator bar (fused silica), and an MCP-PMT with excellent timing resolution, such as the Hamamatsu SL10 \cite{ref7} or Photonis Planacon.  An example of one such radiator bar and readout suitable for the KEKB detector upgrade is shown in Figure \ref{fig:fig1}.  Most geometrical parameters are heavily constrained by the PID detector envelope.  As such, the expansion volume is limited to a distance of order $\sim$5 cm.  The exact size of the image plane is chosen to match the photodetector active area and ensure that light travels from the end of the radiator bar through the expansion region with no further reflections in the vertical direction.

\section{Simulated results}

\begin{figure}[hbt]
\begin{center}
\includegraphics*[scale=0.44]{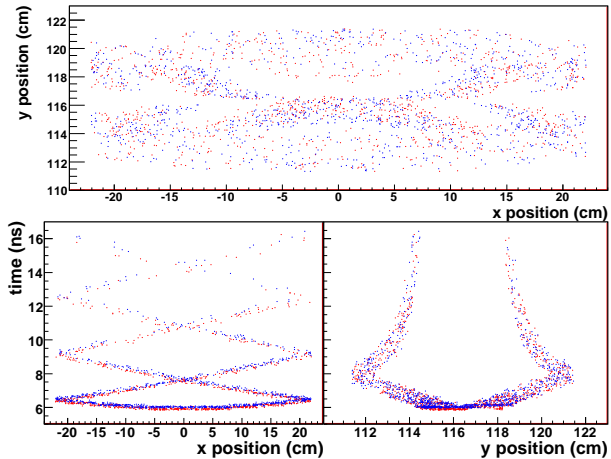}
\end{center}
\caption{\label{fig:fig2a}
Simulated photon hits at an iTOP detector plane placed 36 mm from the end of a 44 cm x 2 cm x 271 cm radiator bar.  The hits correspond to 2 GeV/c charged tracks emitted at a 120 degree polar angle from the interaction point, shown as a function of y position versus x position (top), time versus x position (bottom left), and time versus y position (bottom right). The detected time includes both time-of-flight of the charged particle and time-of-propagation of the photons within the radiator bar.
}
\end{figure}

Geant4 based simulations have been performed to evaluate the potential benefits of an iTOP over a basic TOP.  Geant4 is used to model the emission of Cerenkov radiation within the radiator material, and to propagate the photons to the detection plane.  Example photon patterns in position and time are shown in Figure \ref{fig:fig2a}.  Using a log likelihood difference method, these patterns of photon hits for given track angles and momenta are used to determine kaon efficiencies and pion fake rates, which in turn define the separability of a given geometry.  Effects of the photodetector pixelization, efficiency, and associated timing resolution are added in during these analysis procedures.

%
%
%
%
%

\begin{figure}[hbt]
\begin{center}
\includegraphics*[scale=0.50]{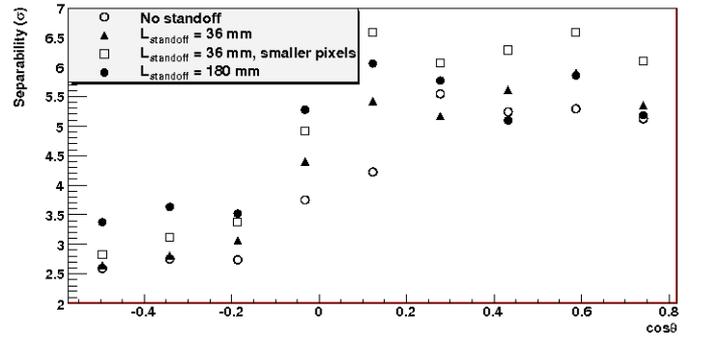}
\end{center}
\caption{\label{fig:fig2}
Simulated $K/\pi$ separability in number of $\sigma$ for a 4 GeV/c charged track as a function of the cosine of track polar angle.  Results are shown for a pure TOP counter (empty circles), an iTOP with a 36 mm expansion length (filled triangles), 180 mm expansion length (filled squares), and a 36 mm expansion length with finer vertical pixelization (open squares).  All results use a 44 cm wide, 2 cm thick, 270 cm length radiator bar, with a focusing mirror in the forward direction.  The focal length is chosen as the sum of the radiator length plus the standoff wedge length.  The photodetector is assumed to have an effective pixel size consistent with the SL10 (22 cm horizontal x 0.55 cm vertical) and a gaussian transit time spread of 40 ps.  In the case of the points noted with finer vertical pixelization, the vertical pixel size is reduced by a factor of 5.
}
\end{figure}

Sample simulated separability results are shown in Figure \ref{fig:fig2}.  There is a clear benefit to a large standoff region, and therefore a large detector plane.  However, as explained in the previous section, the largest practical standoff length that can be accommodated in the barrel is $\sim$5 cm.  In the presence of this size limitation, finer segmentation of the readout plane allows the performance to be further extended.  All geometries currently being studied in simulation utilize a focusing mirror in the forward direction.  All iTOP geometries considered thus far allow for $>4\sigma$ $K/\pi$ separability in the forward direction.  Performance in the backward direction is limited due to the shorter photon propagation length and lack of focusing.

%
%
%
%
%

\section{Engineering design and simulation validation}

Engineering studies of the basic iTOP concept have begun, including support structure for the radiator bars, expansion areas, and readouts.  Other details, such as specific readout electronics and associated data processing are being studied at a cosmic ray test stand at the University of Hawaii \cite{ref10}.  Optimum geometrical configurations are being studied from simulation, and serve to guide further engineering exercises.  The Hawaii test stand will also be used to validate and tune the Geant4 simulations.  Furthermore, refinements in reconstruction algorithms which efficiently utilize the redundancy between the spatial and timing information are being developed, and may further improve the performance of an iTOP device.

\section{Conclusion}

The iTOP is an extension to the basic TOP PID device which maintains the slim detector profile of the TOP counter.  Simulated results indicate that improved performance in the $K/\pi$ separation power is potentially significant with a large standoff region, fine pixelization, or a combination of the two.  Work is ongoing to determine the ideal geometrical configuration for such a device.  Guidance and validation from prototype testing will determine the degree to which the improvements expected from simulated studies may be realized in a final configuration.

\section*{Acknowledgments}

This work has been supported by the Department of Energy Advanced Detector Research Program Award \#DE-FG02-08ER41571.

%
%
%
%
%
%

\end{document}